\documentclass[a4paper,12pt]{article}
\usepackage{graphicx}% Include figure files
\usepackage{dcolumn}% Align table columns on decimal point
\usepackage{bm}% bold math
\usepackage{authblk}
\usepackage{hyperref}
\usepackage{setspace}
\usepackage{authblk}
\usepackage{siunitx}
\newcommand{\be}{\begin{equation}}
\newcommand{\ee}{\end{equation}}

\title{\textbf{\Large The effect of the fluctuating proton size\\ on the study of chiral magnetic effect\\ in proton-nucleus collisions}\\}
\author[1,2]{\normalsize Dmitri Kharzeev}
\author[3]{Zhoudunming Tu}
\author[3]{Aobo Zhang}
\author[3]{Wei Li}
\affil[1]{\textit{\normalsize Department of Physics and Astronomy, Stony Brook University, New York 11794, USA}}
\affil[2]{\textit{\normalsize Department of Physics and RIKEN-BNL Research Center, \newline Brookhaven National Laboratory, Upton, New York 11973, USA}}
\affil[3]{\textit{\normalsize Department of Physics and Astronomy, Rice University, Houston TX 77054, USA}}

\date{}

\usepackage[left=0.95in,top=0.6in,right=0.95in,bottom=0.6in]{geometry} % Document margins

\begin{document}
\maketitle

\begin{abstract}
High energy proton-nucleus (pA) collisions provide an important constraint on the study of 
the chiral magnetic effect in QCD matter. Naively, in pA collisions one expects no correlation between the orientation of event plane as reconstructed from the azimuthal distribution of produced hadrons and the orientation of magnetic field. If this is the case, any charge-dependent hadron correlations can only result from the background.  Nevertheless, in this paper we point out that in high multiplicity pA collisions a correlation between the magnetic field and the event plane can appear. This is because triggering on the  high hadron multiplicity amounts to selecting Fock components of the incident proton with a large number of partons that are expected to have  a  transverse size much larger than the average proton size.  We introduce the effect of the fluctuating proton size in the  Monte Carlo Glauber model and evaluate the resulting correlation between the magnetic field and the second-order event plane in both pA and nucleus-nucleus (AA) collisions. The fluctuating proton size is found to result in a significant correlation between magnetic field and the event plane  in pA collisions, even though the magnitude of the correlation is still much smaller than in AA collisions. This result opens a possibility of studying the chiral magnetic effect in small systems. 

\end{abstract}

\section{\label{sec:intro} Introduction}
Quantum Chromodynamics (QCD) is expected to possess a rich vacuum structure prescribed by the compact non-Abelian gauge group. Even though QCD is known to respect P and CP invariances, the topological transitions in the QCD vacuum induce locally the violation of parity by producing a chirality imbalance between the left- and right-handed quarks. 
Moreover, in the vicinity of the deconfinement phase transition metastable domains with broken parity may emerge ~\cite{Lee:1973iz,Lee:1974ma,Morley1985,Kharzeev:1998kz}. It has been proposed that the local parity violation induced by topological transitions can be detected in heavy ion collisions through the event-by-event fluctuations in the charge asymmetry of produced hadrons relative to the reaction plane \cite{Kharzeev:2004ey}. This asymmetry is caused by the electric current along magnetic field induced by chiral anomaly in the presence of chirality imbalance, or the chiral magnetic effect (CME) \cite{kharzeev2008effects,fukushima2008chiral}, see \cite{kharzeev2014chiral,Kharzeev:2015znc} for reviews. 
It has been proposed to detect the CME by using a charge-dependent three-particle correlator~\cite{Voloshin:2004vk}.

The chiral magnetic current is quenched by the spontaneous breaking of chiral symmetry \cite{kharzeev2008effects,fukushima2008chiral}, and therefore an experimental observation of the CME would also represent a direct evidence for the chiral symmetry restoration~\cite{Kharzeev:2015znc}. In the past decade, the charge-dependent three-particle correlator ($\gamma$-correlator) has been measured in nucleus-nucleus (AA) collisions by the STAR Collaboration at BNL RHIC and ALICE Collaboration at CERN LHC~\cite{Abelev:2009ac,Adamczyk:2014mzf,Adamczyk:2013hsi,Abelev:2012pa,Acharya:2017fau}. The results have been found to be consistent with the CME expectations. However, non-negligible backgrounds related to the elliptic flow, momentum conservation, local charge conservation, and other short-range correlations have been identified and are also qualitatively consistent with the data~\cite{Schlichting:2010qia,Wang:2009kd,Wang:2016iov,Bzdak:2012ia,Bzdak:2010fd}, see \cite{Kharzeev:2015znc} for a review. 

Recently, CMS Collaboration reported a new approach to constraining the CME and background effects using the high-multiplicity proton-lead (pPb) data, where significant charge-dependent signal has been observed with a similar magnitude to that in lead-lead (PbPb) collisions~\cite{Khachatryan:2016got,Sirunyan:2017quh}. These results not only constrain both the CME interpretation and the background models, but also open the possibility to study the CME in high-multiplicity pA collisions. 

The importance of pA collisions for the study of CME stems from the fact that the event plane as determined from the hadron azimuthal distribution is not correlated with the impact parameter of the collision, as the nucleons struck by the proton are randomly distributed within the nucleus ~\cite{Khachatryan:2016got,Sirunyan:2017quh}, see also ~\cite{belmont2016cme,Deng:2016knn} for Monte Carlo studies of this correlation. The absence of the correlation is due to the fact that the proton size is much smaller than the size of the nucleus, and so the incident proton probes the nucleus at small spatial scales. On the other hand, the magnetic field is produced coherently by all protons in the nucleus, and so its orientation is perpendicular to the impact parameter of a pA collision. Since the CME is driven by magnetic field, and the background effects -- by the hadron event plane, one can use the measured 
$\gamma$-correlator to constrain the magnitude of CME. 

However, while it seems safe to assume that the proton size is small $R_p \simeq 1$ fm, this may not be so in high multiplicity pA collisions studied in the CMS experiment  ~\cite{Khachatryan:2016got,Sirunyan:2017quh}. Indeed, at high energies the Fock states of the proton's wave function with different numbers of partons are frozen due to the time dilation, and interact with the target with different probabilities. This picture is the basis of the theory of hadron diffraction \cite{feinberg1956nuovo,good1960diffraction}, and underlies the Glauber-Gribov theory of inelastic shadowing \cite{Gribov:1968jf}. Moreover, the observed shrinkage of diffraction cone in proton-proton scattering indicates the growth of the proton size with energy, when the number of partons in the proton's wave function increases (the ``Gribov diffusion" \cite{Gribov:2009zz}):
\begin{equation}
\label{radius_1}\label{gribov_diff}
R^{2}_{p}(s) \approx R^{2}_{p}(0) + 2 \alpha^{'} \ln{s}
\end{equation}
where the $R_{p}(0)$ is the proton radius in the rest frame, $\alpha^{'} \simeq 0.25\ {\rm GeV}^{-2}$ is the slope of the Pomeron trajectory, and $s$ is the square of the center-of-mass energy.
The recent data \cite{Antchev:2016vpy,Csorgo:2016qyr,Berretti:2017fhy} from TOTEM and CMS collaborations at the LHC show that the growth of the proton size becomes even faster, that has been attributed to higher order string effects \cite{Kharzeev:2017azf}. The effects of the proton size fluctuations on the attenuation of nuclear cross sections and the number of wounded nucleons have been found quite significant, see \cite{Brodsky:1988xz,Alvioli:2013vk,Kopeliovich:2016jjx} and references therein. 

The growth of the proton size with energy reflects the increase in the number of partons that is governed by the QCD evolution. However there is an alternative, and much more efficient, way to select the proton configurations with a large number of partons, and thus with a large size. Namely, one can trigger on the events with a high multiplicity. Indeed, it has been found recently that the distribution in the number of partons predicted by QCD evolution equations describes quite well the measured multiplicity distributions in high energy proton-proton collisions \cite{Kharzeev:2017qzs}. This suggests that the ``local parton-hadron duality" \cite{Dokshitzer:1987nm} holds also on the event-by-event basis, and triggering on the events with high hadron multiplicity one selects the proton configurations with a large number of partons and thus a larger than average size. 

\begin{figure*}[thb]
\centering
\includegraphics[width=3.4in]{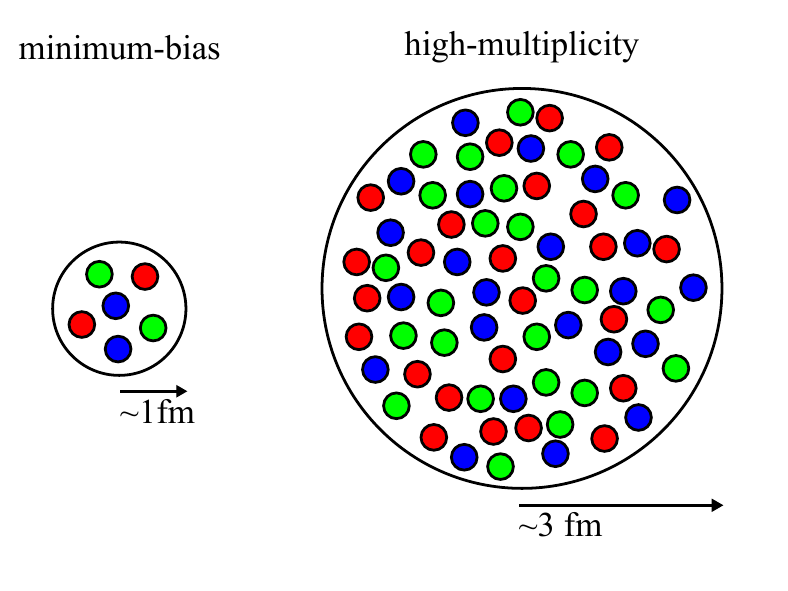}
  \caption{ \label{fig:cartoon} 
 A sketch of the proton configurations probed in the minimum bias (left) and high multiplicity (right) events. }
\end{figure*}

In the Gribov diffusion picture  \cite{Gribov:2009zz}, each splitting in the parton ladder represents a step in the random walk in the transverse plane, with the transverse size squared growing proportionally to the number of splittings. Rapidity $y = \ln (s/s_0)$ plays the role of time in this diffusion process, which immediately leads to (\ref{gribov_diff}). The same picture applied to the events with high hadron multiplicity $N$ makes us conclude that the proton configuration size $R_p$ grows according to
\begin{equation}
\label{radius_2}\label{prot_size}
R^{2}_{p}(N) \approx \frac{N}{\bar{N}}R^{2}_{0} ,
\end{equation}
where $R_0$ is the average size corresponding to the event-averaged hadron multiplicity  $\bar{N}$. The relation (\ref{prot_size}) allows to estimate the size of the proton configurations, and in high-multiplicity pA collisions one selects events with a large proton size, as illustrated in Fig.~\ref{fig:cartoon}. For example, selecting events with a multiplicity that is ten times larger than average, we trigger on the proton configuration size of over 3 fm corresponding to the size of  a light nucleus with mass number $A \simeq 25$.

It is clear that the large size of the proton configuration will affect the correlation between the event plane and the impact parameter. This correlation is crucial for the study of CME, since 
the CME signal is expected to be proportional to the magnitude of the magnetic field and its correlation with the second-order event plane~\cite{Deng:2016knn}, which can be expressed as

\begin{equation}
\label{gamma_1}
\Delta\gamma \propto \left\langle |B^{2}| \cos{2(\Psi_{\mathrm B}-\Psi_{\mathrm{EP}})} \right\rangle
\end{equation}

\noindent where $B$ denotes the magnitude of the magnetic field, $\Psi_{\mathrm B}$ and $\Psi_{\mathrm{EP}}$ represent the plane defined by the magnetic field direction and the second-order anisotropy, respectively. In a high-multiplicity pA collision with a fixed-size proton, it has been shown that the correlation between $\Psi_{\mathrm B}$ and $\Psi_{\mathrm{EP}}$ is very small comparing to that in AA collisions~\cite{Khachatryan:2016got}. In this paper we repeat this analysis taking into account the fluctuations in the proton size, using a Monte Carlo (MC) Glauber approach~\cite{Alver:2008aq}, and explore the implications for the studies of CME in pA and AA collisions.
\vskip0.3cm
The paper is organized as follows. In Section~\ref{sec::theory}, we introduce the idea of correlation between the final-state particle multiplicity and the initial-state proton size. In Section~\ref{sec:mc}, we summarize the setup of MC Glauber model with a fluctuating proton size. In Section~\ref{sec:results}, the results are presented in terms of the correlation that is sensitive to the CME observable, and in Section~\ref{sec:summary}, we summarize the study and discuss its implications for the studies of CME in pA and AA collisions.

\section{\label{sec::theory} Fluctuating proton size: theory expectations}

Hadrons are relativistic bound states of strongly interacting quarks and gluons. They represent 
eigenstates $| n \rangle$ of the QCD Hamiltonian $H_{\rm QCD}$ with a definite mass $M_n$:
\be
H_{\rm QCD} | n \rangle = M_n | n \rangle .
\ee
Different hadron states with the same quantum number are separated in mass by $\Delta M \sim \Lambda_{QCD} \sim {\cal{O}} (200\ {\rm MeV})$. %For example, the excited state of the proton is the Roper resonance with the mass of 1440 MeV.

\vskip0.3cm
Let us now consider a high-energy proton-proton collision in the center-of-mass frame (the laboratory frame of the collider experiments). In this frame, the colliding protons are Lorentz-contracted to thin pancakes of thickness $R_{\rm QCD}/\gamma \sim (\gamma\ \Lambda_{QCD})^{-1}$, where the gamma-factor is $\gamma = \sqrt{s}/(2 M)$ ($s$ is the square of the c.m.s. energy, and  $M$ is the proton mass). The collision thus occurs over the short time interval 
\be\label{colltime}
\Delta t \sim \frac{R_{\rm QCD}}{\gamma} \sim \frac{1}{\gamma\ \Lambda_{QCD}} \ll \frac{1}{\Delta M}.
\ee 
The uncertainty principle dictates that this collision is not able to resolve the individual mass eigenstates, and so highly excited states with masses extending all the way to $\sim \sqrt{s}/2$ can contribute to the collision. In other words, a high energy proton-proton collision cannot be described in terms of the lowest mass eigenstates of the QCD Hamiltonian with the quantum numbers of the proton. 
\vskip0.3cm

At high energies, a convenient description is in terms of the Fock basis that contains the states with a fixed number of partons. The positions of these partons are ``frozen" since during the short collision time (\ref{colltime}) they cannot move in the transverse plane, or split into a different parton configuration. At large momentum transfer $Q^2$, the evolution of parton densities with Bjorken $x$ and $Q^2$ can be described by perturbative QCD evolution equations. At high enough energy, the parton densities inside the protons saturate, and can be characterized by a saturation momentum $Q_s(x)$ \cite{Gribov:1984tu,McLerran:1993ni}. However the proton size is determined by partons that are close to the edge of the proton, where the parton density is never large. In this domain, the dynamics of parton splitting is non-perturbative, and has to be described by an effective theory that incorporates confinement. 
\vskip0.3cm

Such an effective description pre-dates QCD and is known as the Reggeon field theory. In this approach, the spectrum of hadrons is described by the linear Regge trajectories $\alpha(t) = \alpha(0) + \alpha^\prime t$, where $t = M^2$ at $t>0$; continuation to negative $t<0$ allows to describe the scattering amplitudes at momentum transfer $t$. The slope $\alpha^\prime$ and the intercept $\alpha(0)$ are determined by the non-perturbative QCD dynamics. The contributions of Regge trajectories to the total interaction cross section are determined by the corresponding intercepts, 
$\sigma_{tot} \sim s^{\alpha(0)-1}$, and so at high energies the dominant contribution arises from the Pomeron trajectory with the intercept $\alpha_P(0)\simeq 1.09$. The profile of the scattering amplitude $A(s; b)$ in the impact parameter plane broadens as a function of rapidity $y = \ln (s/s_0)$:
\be\label{ampl}
A(s; b) \sim \exp\left(-\frac{b^2}{4 y \alpha_P^\prime}\right) ,
\ee
where the slope of the Pomeron trajectory is $\alpha_P^\prime \simeq 0.25\ {\rm GeV}^{-2}$.
This  leads to the increase of the slope $B$ of elastic cross section at high energies:
\be\label{slope} 
B_{el}(s) = B_0 + 2 \alpha_P^\prime \ln s ,
\ee
that is well established and continues up to the highest LHC energy \cite{Antchev:2016vpy,Csorgo:2016qyr,Berretti:2017fhy}. 
\vskip0.3cm

The relations (\ref{ampl}) and (\ref{slope}) imply the linear growth of the proton size squared with rapidity $y = \ln (s/s_0)$, as given by (\ref{gribov_diff}).
Within the Reggeon field theory developed by Gribov, this growth is a consequence of parton diffusion in the transverse plane \cite{Gribov:2009zz}; rapidity plays the role of time in this diffusion process, and each step corresponds to the parton splitting. The average size of the proton  is thus determined by the average number of partons generated by the QCD evolution at a given rapidity.
\vskip0.3cm

Let us now move to the discussion of fluctuations in the proton size. According to the arguments presented above, these fluctuations should be determined by the fluctuations in the number of partons at a given rapidity. In experiment, we have access to the fluctuations in the number of produced hadrons; are these fluctuations related to the distributions in the number of partons within the colliding protons? Recently, it has been observed \cite{Kharzeev:2017qzs} that the distribution in the number of partons predicted by QCD evolution describes quite well the distribution in the number of produced hadrons measured by the CMS Collaboration at the LHC \cite{Khachatryan:2010nk}. This observation is supported by theoretical arguments on the conversion of entanglement entropy within the colliding hadrons into the Boltzmann entropy in the final multiparticle state \cite{Kharzeev:2017qzs} during a rapid ``quench" induced by the collision. 
\vskip0.3cm

Even though the understanding of multiparticle production is far from complete, the arguments presented above suggest that the high multiplicity events originate from the protons with a large  number of partons. Basing on the similarity between the parton and hadron multiplicity distributions, we thus assume that the size of the proton $R_{p}(N)$ probed in a collision producing $N$ hadrons is related to the average size of the proton $R_{0}$ by the relation (\ref{prot_size}). 
 This relation implies that triggering on high-multiplicity collisions we select events in which the incident proton has a size that is significantly larger than an average proton size. For example, assuming that we select events with $N/\bar{N} \simeq 10$, and that the average size of the proton at LHC energy is $R_0 \simeq 1\ {\rm fm}$, the size of the selected proton configurations is over $3$ fm. 
This large size significantly affects the geometry of the proton-nucleus collision, and the correlation between the reaction plane and the magnetic field. We now proceed to the discussion of these effects.

%\clearpage

\section{\label{sec:mc}Monte Carlo Glauber implementation}
\begin{figure*}[thb]
\centering
\includegraphics[width=2.7in]{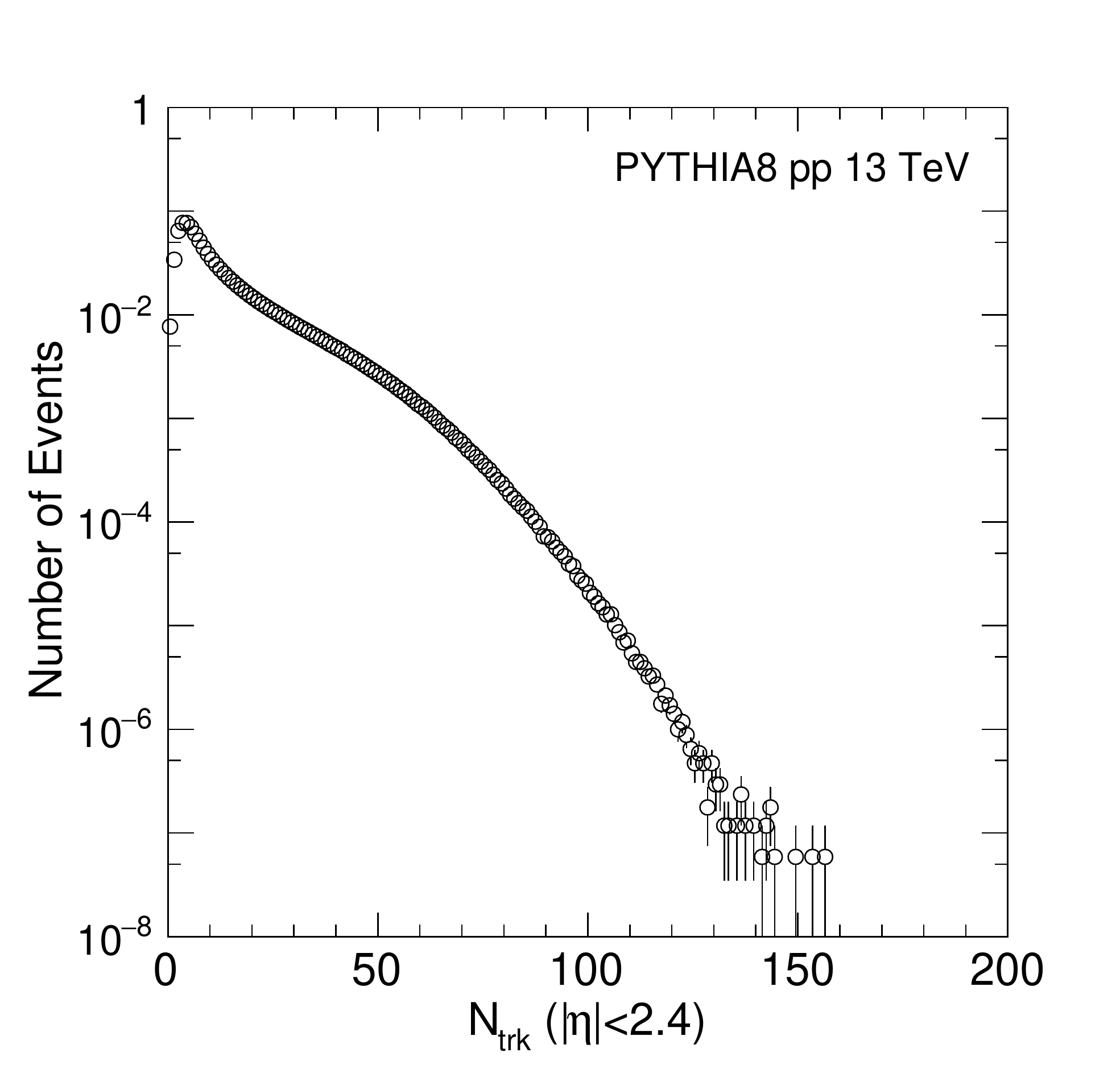}
\includegraphics[width=2.7in]{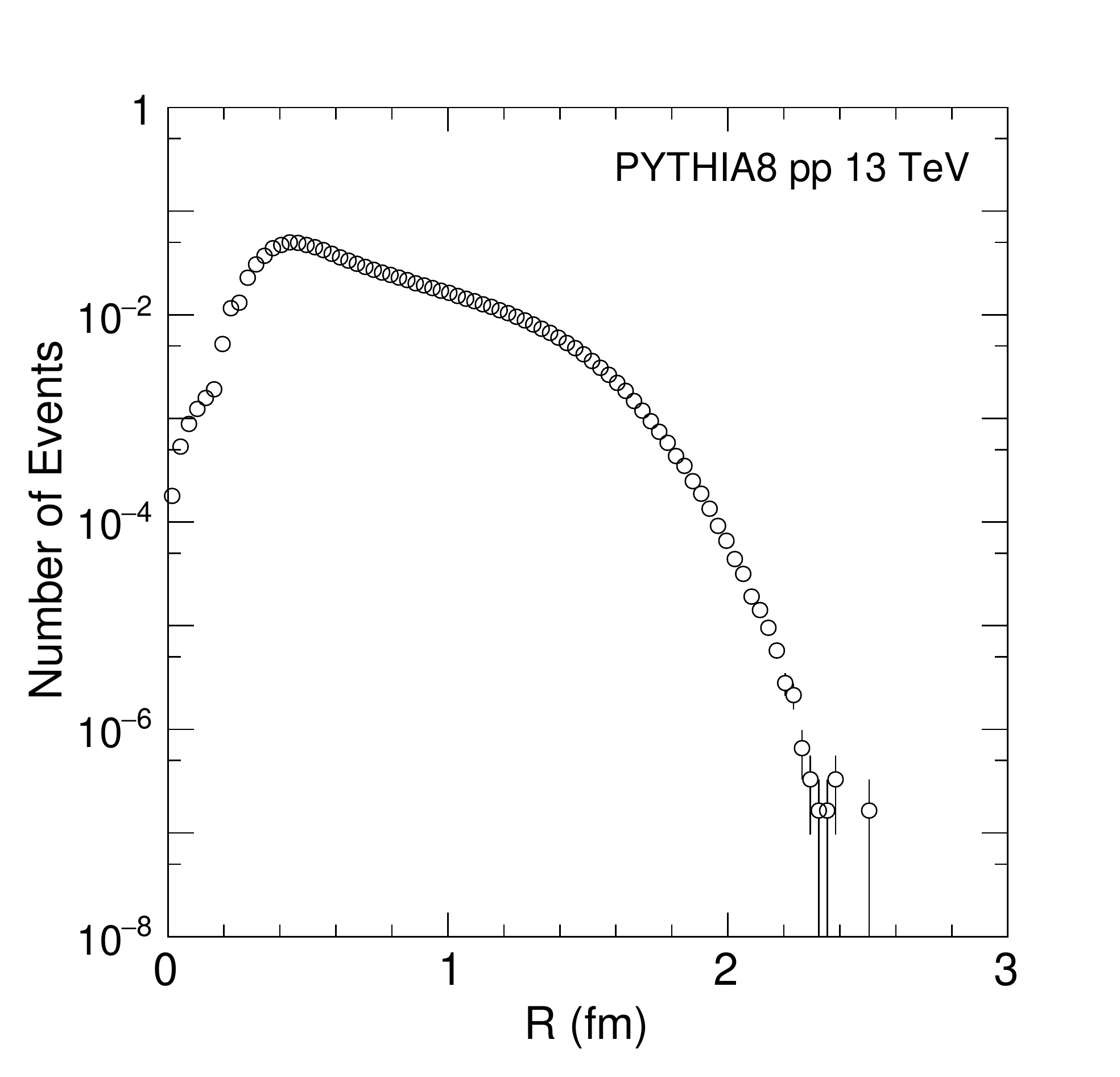}
  \caption{ \label{fig:figure_1} Left: charged-particle multiplicity distribution in pp collision at $\sqrt{\mathrm s}=13$ TeV from \textsc{PYTHIA 8} event generator is shown. Right: the fluctuating proton radius is shown.  }
\end{figure*}

As given by Eq.~\ref{radius_2}, the event-by-event proton configuration transverse size is proportional to the square root of the multiplicity $\sqrt{N}$ in each event. In order to obtain a realistic fluctuating proton size, the charged-particle multiplicity distribution 
($N_{\rm trk}$) is taken from a \textsc{PYTHIA 8} MC sample at the center-of-mass energy at $\sqrt{\mathrm s}=13$ TeV, where $N_{\rm trk}$
is counted within the kinematic coverage of $p_{T} > 0.4$ GeV/c and $|\eta|<2.4$, same to that used by the CMS collaboration~\cite{Khachatryan:2010gv,Khachatryan:2012dih,Chatrchyan:2013nka, Khachatryan:2016yru, Khachatryan:2016txc, Khachatryan:2016got}. According to the measured average 
cross section for inelastic pp collisions at $\sqrt{\mathrm s}=13$ TeV~\cite{Aaboud:2016mmw}, $\sigma^{pp}_{in} \approx 70$ \si{mb}, the average proton radius can be translated into $R_{\mathrm 0} \approx 0.75$ \si{fm}, if an inelastic collision takes place when the distance between two protons is less than twice of their radius. In Fig.~\ref{fig:figure_1}, the charged-particle multiplicity distribution at $\sqrt{\mathrm s}=13$ TeV (left) from the \textsc{PYTHIA 8} event generator and the resulting fluctuating proton radius (right) from Eq.~\ref{radius_2} are presented. As one can see, the proton radius is no longer a fixed value but fluctuates event-by-event, and the radius can be as large as 2.5 \si{fm}.

Instead of using a fixed proton radius of 0.75 \si{fm}, the distribution of the fluctuating proton radius is used as the input in the MC Glauber model for the projectile proton. The nucleons in the nuclei are sampled with a fixed radius of 0.75 \si{fm}, as the projectile proton strikes several nucleons and the size of each nucleon fluctuates independently, thus leading to negligible effects on average.
The criterion for the wounded nucleon requires the distance between the two nucleons to be less than the sum of their radii, similar to the default MC Glauber model. The lead nucleus with 208 nucleons and a radius of 6.62 \si{fm} for the Woods-Saxon distribution is used for this study, and no modification of the MC Glauber model is applied for the PbPb collisions. 

With this setup, the impact parameter of the collision ($b$), number of participants ($N_{\mathrm part}$), the participant plane ($\Psi_{\mathrm{PP}}$), the reaction plane ($\Psi_{\mathrm{RP}}$), and eccentricity can be calculated event-by-event. The magnetic field direction is approximated by the direction that is perpendicular to $b$, and its magnitude is estimated to be proportional to the magnitude of the impact parameter ($B \propto b$). The $\Psi_{\mathrm{PP}}$ is used to approximate the second-order event plane, $\Psi_{\mathrm{EP}}$, of final-state hadrons. Therefore, the correlation between the $\Psi_{\mathrm B}$ and $\Psi_{\mathrm{EP}}$ can be studied using $\Psi_{\mathrm{RP}}$ and $\Psi_{\mathrm{PP}}$, 
which is related to the charge-dependent correlator $\Delta\gamma$ from the CME.

\section{\label{sec:results}Results}

\begin{figure*}[thb]
\centering
\includegraphics[width=2.5in]{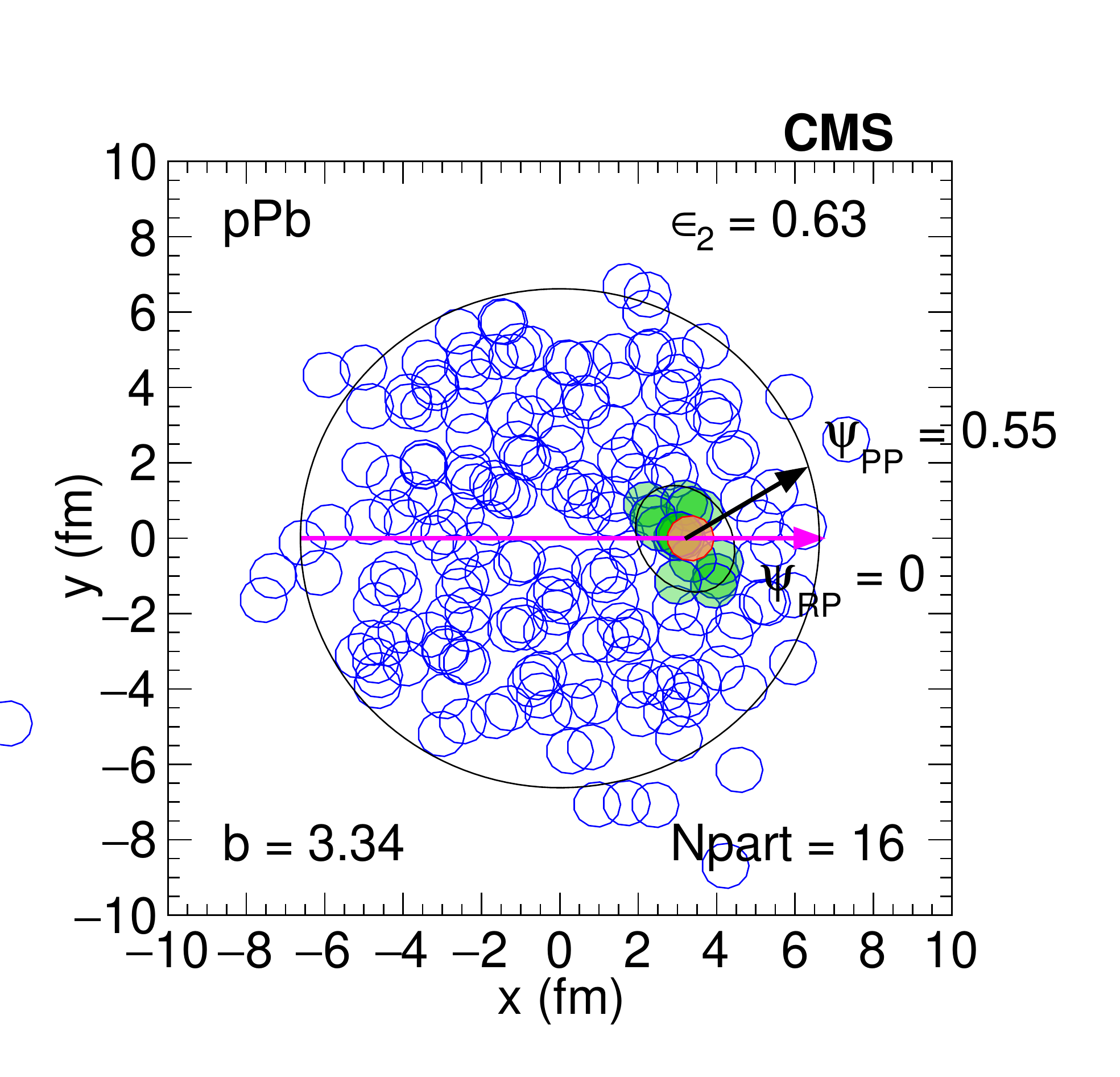}
\includegraphics[width=3.7in]{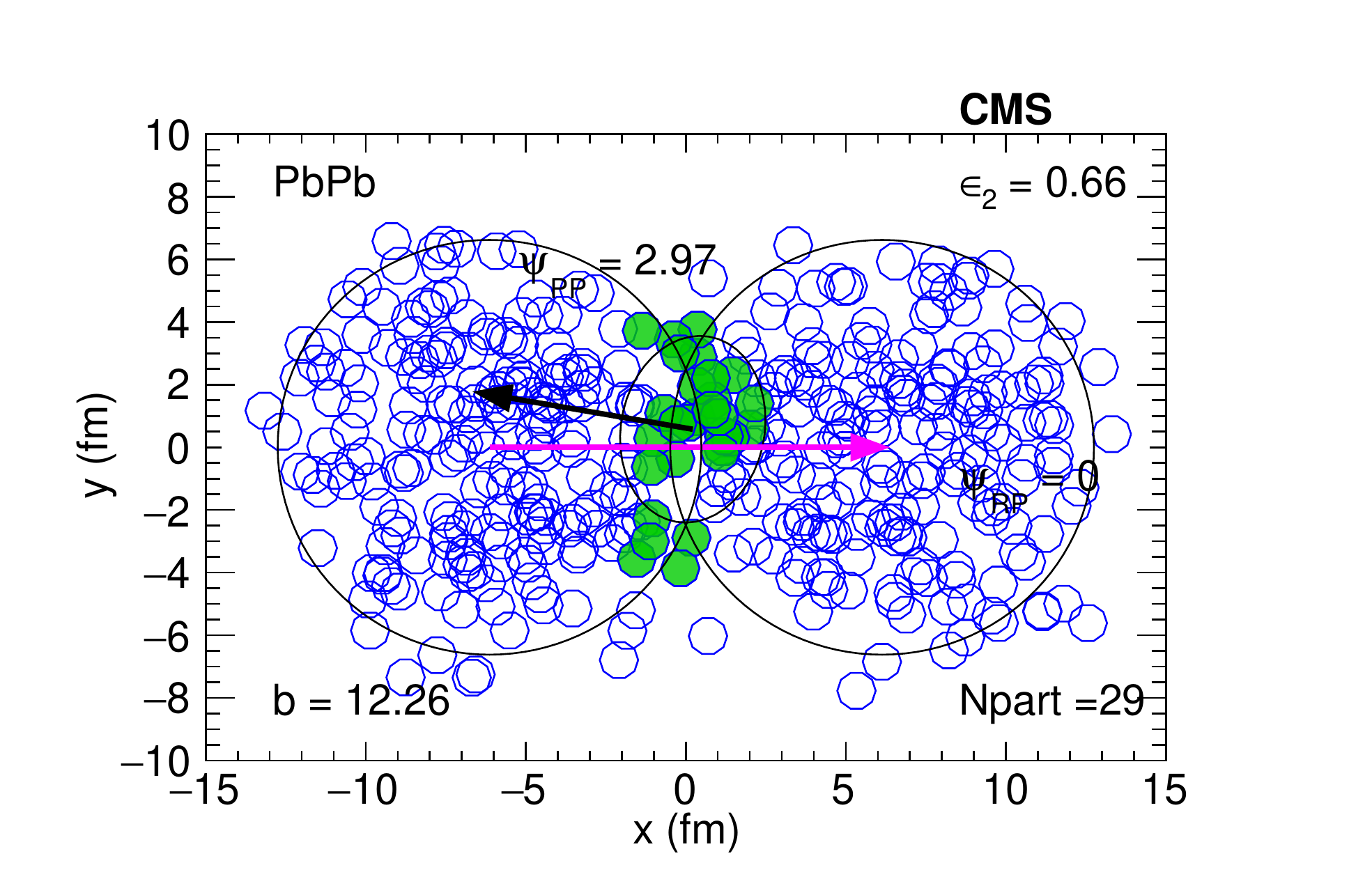}
  \caption{ \label{fig:figure_2} Event displays of one high-multiplicity pPb (left) and a peripheral PbPb event (right) using MC Glauber simulation at $\sqrt{s_{_{\mathrm {NN}}}} = 5.02$ TeV are shown. The red circle is the proton projectile, the green circles are the participant nucleons, and the blue circles are the spectator nucleons. The red and black arrows refer to the angle of the reaction plane ($\Psi_{\rm RP}$) and the participant plane ($\Psi_{\rm PP}$) in the transverse direction~\cite{Khachatryan:2016got}.}
\end{figure*}

Figure.~\ref{fig:figure_2}, published in Ref.~\cite{Khachatryan:2016got} by the CMS Collaboration, shows an event display for one high-multiplicity (large $N_{\mathrm part}$) pPb and a peripheral PbPb event at $\sqrt{s_{_{\mathrm {NN}}}} = 5.02$ TeV using the MC Glauber model simulation. The projectile proton in pPb collision has a transverse radius of 0.75 \si{fm}. The red, green, and blue circles represent the projectile proton, wounded nucleons, and spectator nucleons, respectively. The pink and black arrow denote the angle of the reaction plane $\Psi_{\rm RP}$ and the participant plane $\Psi_{\rm PP}$. Using these angles, the correlation between $\Psi_{\rm B}$ and $\Psi_{\rm EP}$ can be evaluated event-by-event. Therefore, a comparison between pPb and PbPb collisions in terms of their expected magnitude of the $\Delta\gamma$ correlator from the CME signal using Eq.~\ref{gamma_1} can be obtained. 

\begin{figure*}[thb]
\centering
\includegraphics[width=2.7in]{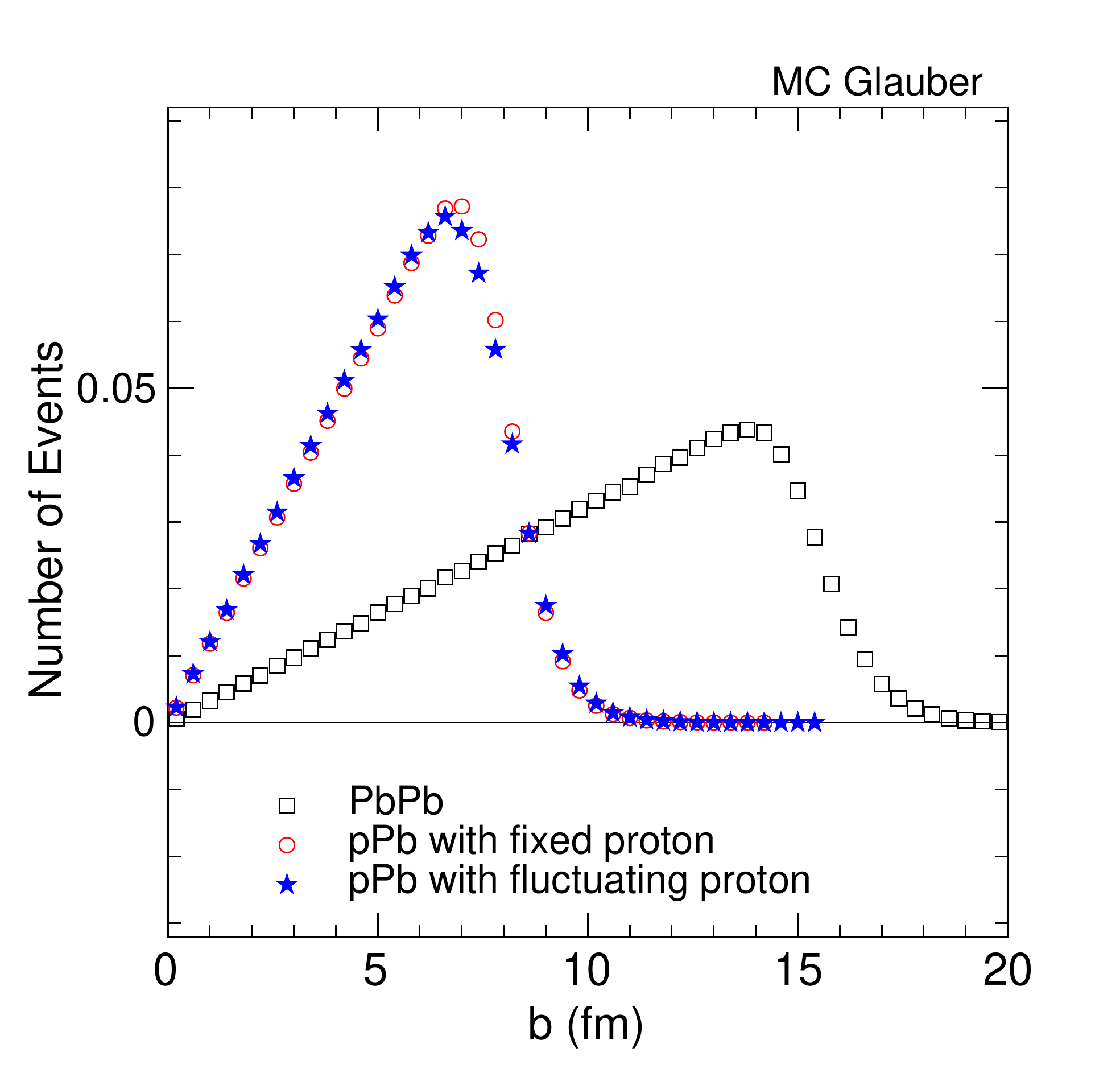}
\includegraphics[width=2.7in]{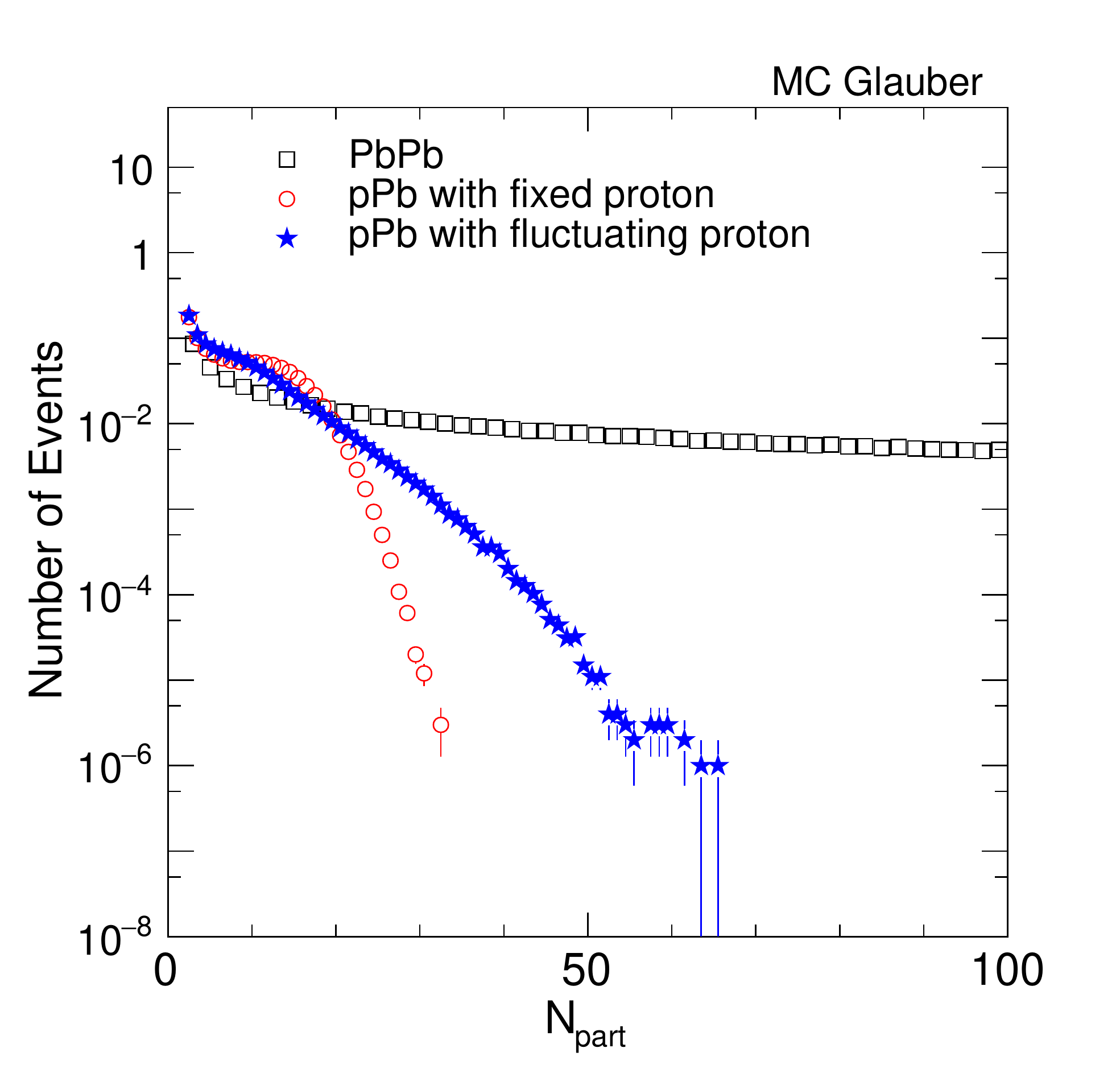}
  \caption{ \label{fig:figure_3} The impact parameter (left) and the number of participants  (right) distributions in PbPb and pPb collisions with fixed and fluctuating proton size as computed within the MC Glauber model with one million simulated events. The statistical uncertainty is shown by the error bar.}
\end{figure*}

Figure.~\ref{fig:figure_3} shows the impact parameter $b$ (left) and the number of participants $N_{\mathrm part}$ (right) distributions in PbPb collisions (black), pPb collisions with fixed (red), and fluctuating proton size (blue) using MC Glauber model with one million simulated events. The impact parameter distribution in PbPb collisions covers a wider range than that in pPb collisions, as expected from geometry of the collision. In pPb collisions, the impact parameter distributions corresponding to the fixed and fluctuating proton size are similar, and both fall between 6--7 \si{fm}, which is expected and consistent with the size of the lead nucleus. In these simulations, the impact parameter is sampled randomly in a given range without any dependence on the size of the projectile proton. In Fig.~\ref{fig:figure_3} (right), the $N_{\mathrm part}$ distribution is also shown with the same three scenarios that have been studied in the left panel. The $N_{\mathrm part}$ can go up to 416 in PbPb collisions but is only shown up to 100 in order to better compare with pPb collisions. In pPb collisions, the $N_{\mathrm part}$ distributions with fixed and fluctuating proton sizes behave very differently after $N_{\mathrm part} \approx 20$, with fluctuating proton leading to a much wider distribution extending to a higher $N_{\mathrm part}$. The fatter proton is more likely to generate a collision with a larger $N_{\mathrm part}$, which is consistent with a less steep fall of the  distribution shown in Fig.~\ref{fig:figure_3} (right). 

\begin{figure*}[thb]
\centering
\includegraphics[width=2.7in]{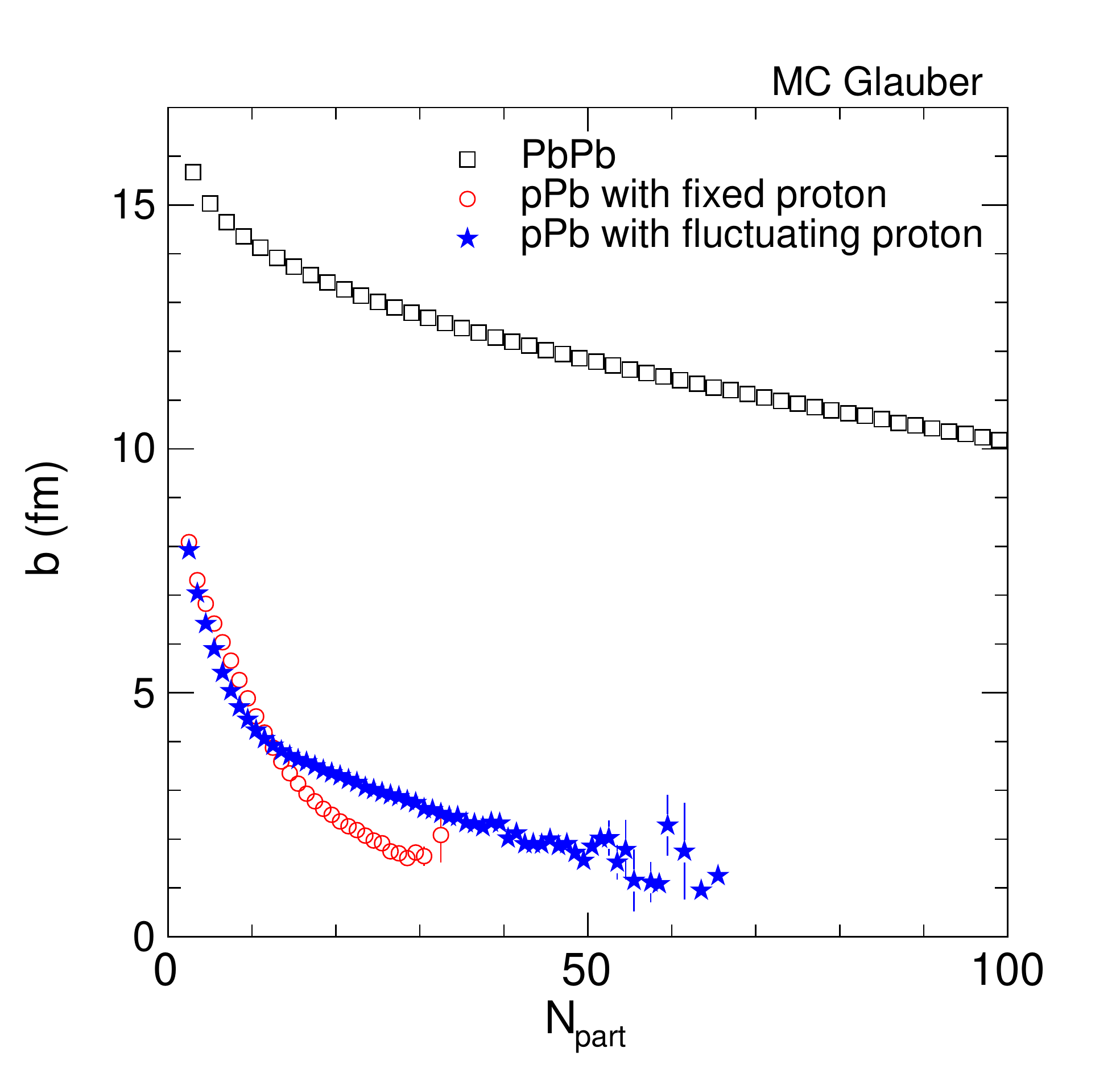}
\includegraphics[width=2.7in]{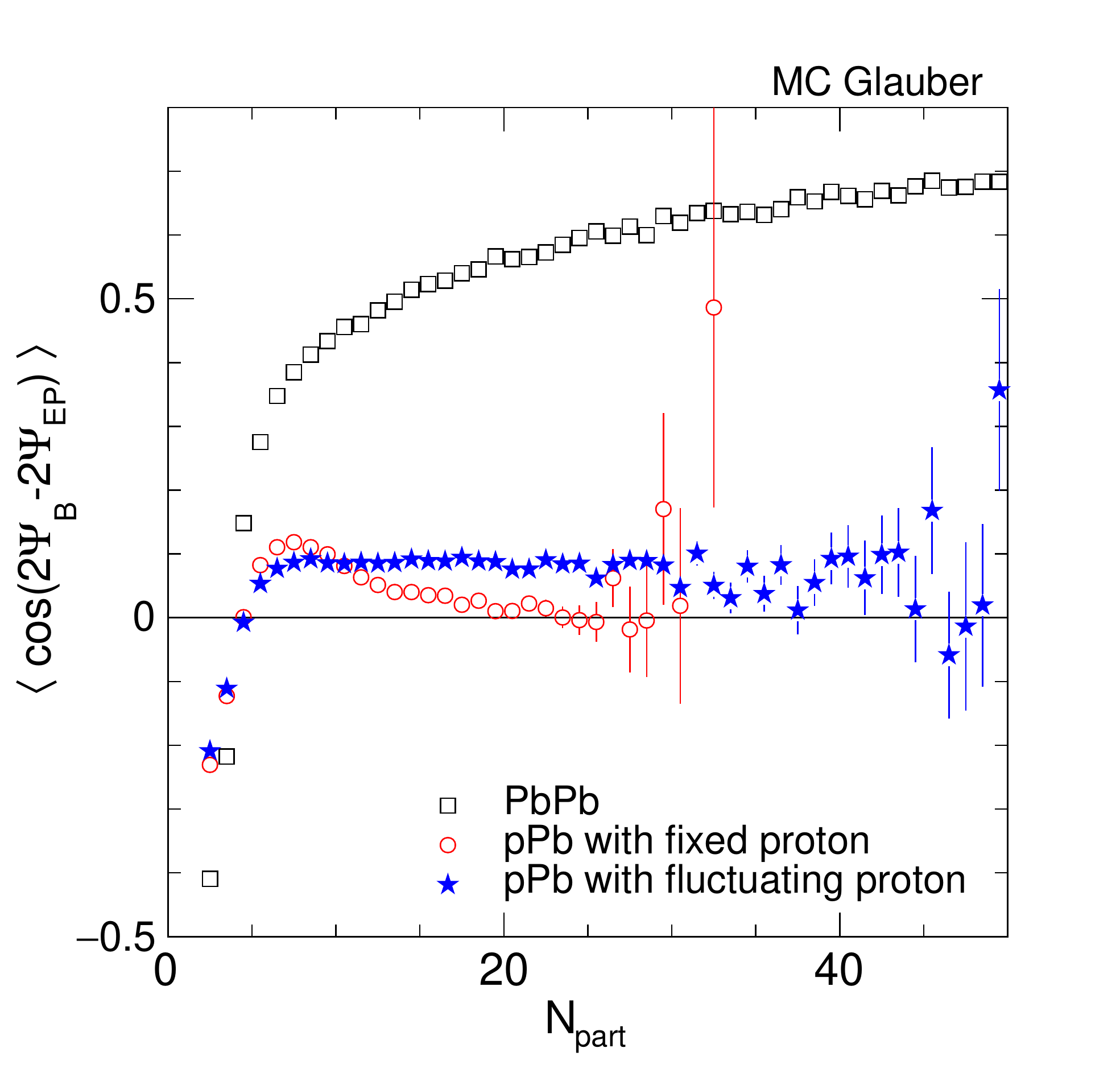}
  \caption{ \label{fig:figure_4} The event-averaged impact parameter (left) and angular correlation (right) as a function of the number of participants in PbPb collisions and pPb collisions with fixed and fluctuating proton sizes, using MC Glauber model with one million simulated events. The statistical uncertainty is shown by the error bar.}
\end{figure*}

In Fig.~\ref{fig:figure_4}, the averaged $b$ (left) and angular correlations, $\left\langle \cos(2\Psi_{B}-2\Psi_{EP}) \right\rangle$ (right), as a function of $N_{\mathrm part}$ are shown in PbPb collisions (black), pPb collisions with fixed (red) and fluctuating proton size (blue), using MC Glauber model simulation with one million events. Qualitatively,  a decreasing impact parameter $b$ results in the increase in $N_{\mathrm part}$, reflecting the fact that more central collisions have a larger number of wounded nucleons, and therefore a larger final-state particle multiplicity. At around $N_{\mathrm part} \approx 10$, the two cases with fixed and fluctuating proton start to deviate from each other. The fluctuating proton scenario has been found to have a larger $N_{\mathrm part}$ at the same $b$, which is consistent with the expectation for a fatter proton. The angular correlation between $\Psi_{\mathrm B}$ and $\Psi_{\mathrm{EP}}$ also shows a different behavior above the same $N_{\mathrm part} \approx 10$:  the fluctuating proton is found to yield a significant nonzero correlation as $N_{\mathrm part}$ increases whereas in the fixed proton scenario the correlation quickly drops to zero. 
Comparing pPb and PbPb systems at same $N_{\mathrm part}$ values, the correlations in fluctuating proton scenario for pPb collisions
are still several times smaller than those expected in PbPb collisions.

\begin{figure*}[thb]
\centering
\includegraphics[width=4.0in]{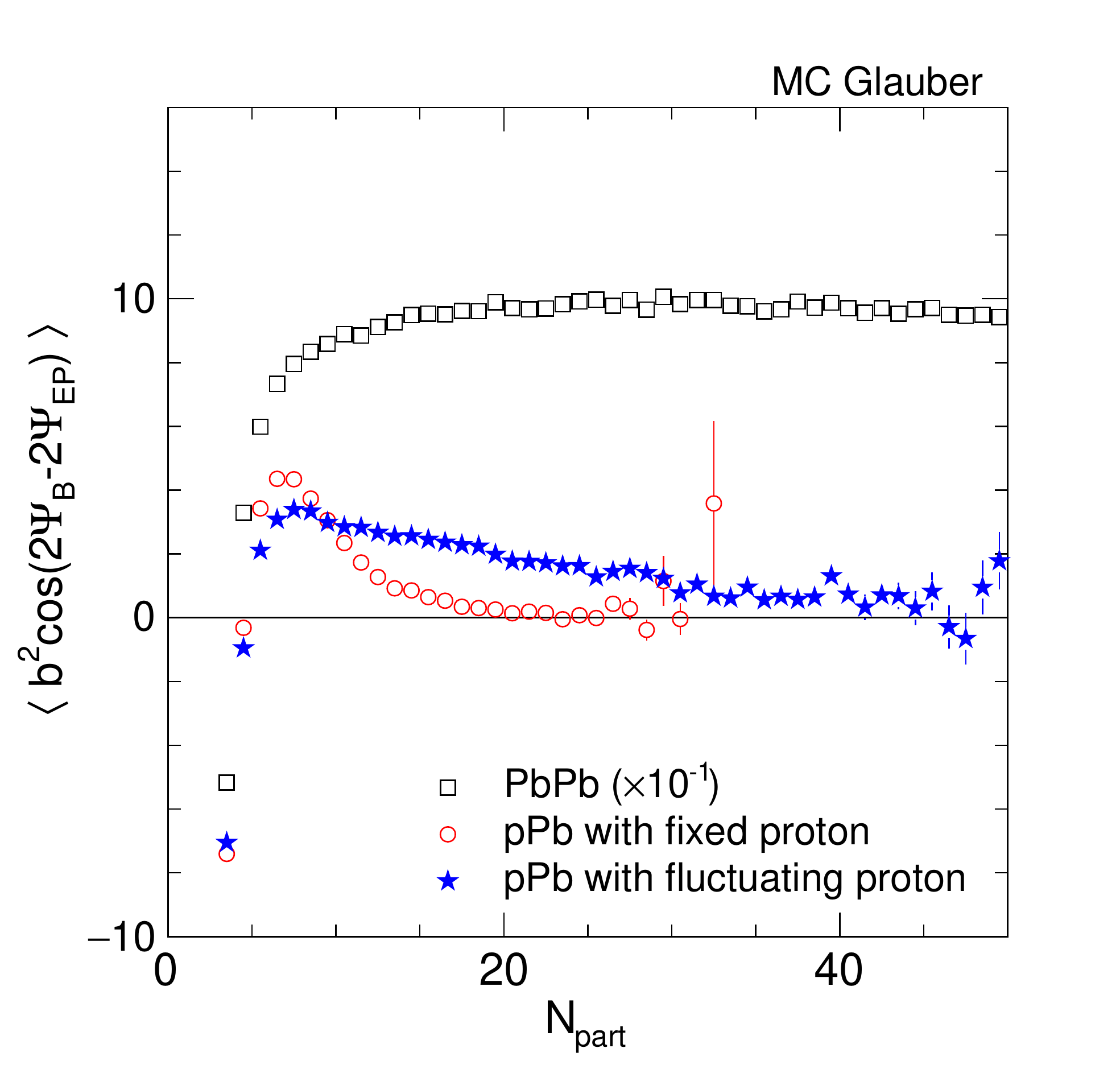}
  \caption{ \label{fig:figure_5} The correlator $\Delta\gamma$ as a function of the number of participants in PbPb collisions and pPb collisions with fixed and fluctuating proton sizes as computed in the MC Glauber model with one million simulated events. For visibility, the PbPb collision data points are scaled by $0.1$ in order to be compared with pPb data. The statistical uncertainty is shown by the error bar. }
\end{figure*}

In order to estimate the expected CME signal, the magnitude of magnetic field needs to be considered as well. For the estimate, we assume that the magnitude of magnetic field is proportional to the impact parameter $b$ (this assumption approximately holds at centralities that we consider \cite{kharzeev2008effects}). Therefore, the correlator $\Delta\gamma$ can be calculated using the impact parameter $b$ and the angular correlation between the magnetic field and the event planes, $\left\langle b^{2}\cos(2\Psi_{B}-2\Psi_{EP}) \right\rangle$, shown in Fig.~\ref{fig:figure_5}. For better visibility, the PbPb data points are scaled by $10^{-1}$ in order to be compared with pPb data. The overall magnitude of the $\Delta\gamma$ correlation in pPb collisions is seen to be enhanced by one order of magnitude within the fluctuating proton scenario. However, the correlation remains significantly smaller than in PbPb collisions.

\clearpage

\section{\label{sec:summary}Summary and discussion}
We have investigated the consequences of the fluctuating proton size for the study of the chiral magnetic effect (CME) in proton-nucleus collisions by using the Monte Carlo Glauber simulation. Due to the initial-state fluctuation of the number of partons, the transverse size of the projectile proton could be much larger than average in high multiplicity collisions.  Using a \textsc{PYTHIA 8} event generator, we generated the proton radius according to the event-by-event particle multiplicity and its event average, and used it as an input distribution of the proton radius in a Glauber simulation. We studied the impact parameter ($b)$, number of participants ($N_{\mathrm part}$), angular correlation between the magnetic field and the event plane, and the expected CME signal correlation in both pPb and PbPb collisions using one million simulated events. The fluctuating proton scenario has been found to yield a much larger $N_{\mathrm part}$, angular correlation, and CME signal with respect to the fixed proton scenario. However, even with the one order of magnitude of enhancement of the signal, this is still much less than that in PbPb collisions. 

This study, for the first time, suggests a possibility of searching for the CME in proton-nucleus collisions. If the presence of CME in nucleus-nucleus collision is established, the search for this effect in high-multiplicity pA collisions could become possible with a large statistics data sample. Given the necessary conditions for the occurrence of CME, the observation of this effect can be a direct evidence of deconfinement and chiral symmetry restoration in small colliding systems. 
\vskip0.3cm

The work of D.K. was supported by the US Department of Energy under contracts No. DE-FG-88ER40388 and DE-AC02-98CH10886. 
The work of Z.T., A.Z. and W.L. was partially supported by an Early Career Award (Contract No. DE-SC0012185) from the US Department of Energy Office of Science, 
the Robert Welch Foundation (Grant No. C-1845) and an Alfred P. Sloan Research Fellowship (No. FR-2015-65911).

\clearpage

\bibliographystyle{elsarticle-num}
\bibliography{reference}

\end{document}